\documentclass[12pt]{article}

\usepackage{amsmath, amssymb, bm}
\usepackage[margin=0.75in]{geometry}
\usepackage{algorithm, algorithmic}
\usepackage{graphicx}
\usepackage{enumerate}
\usepackage{graphicx}
\usepackage{hyperref}

\usepackage{multirow}

\usepackage{float}
\usepackage{lipsum}
\DeclareGraphicsExtensions{.pdf,.png,.jpg}
\usepackage{subcaption}
\allowdisplaybreaks
\usepackage[round]{natbib}
\makeatletter
\def\myunderbar#1{\underline{\sbox\tw@{$#1$}\dp\tw@\z@\box\tw@}}
\makeatother
 \usepackage{siunitx}
\usepackage{setspace}
\usepackage{longtable}
\usepackage{cleveref}
\usepackage{gensymb}
\usepackage{mathtools}

\def\myunderbar#1{\underline{\sbox\tw@{$#1$}\dp\tw@\z@\box\tw@}}

\begin{document}

\title{Compositional Outcomes and Environmental Mixtures: \\the Dirichlet Bayesian Weighted Quantile Sum Regression}
\author{Hachem Saddiki, Joshua L. Warren, Corina Lesseur, Elena Colicino}

\maketitle

\begin{abstract}
Environmental mixture approaches do not accommodate compositional outcomes, consisting of vectors constrained onto the unit simplex. This limitation poses challenges in effectively evaluating the associations between multiple concurrent environmental exposures and their respective impacts on this type of outcomes. As a result, there is a pressing need for the development of analytical methods that can more accurately assess the complexity of these relationships.
Here, we extend the Bayesian weighted quantile sum regression (BWQS) framework for jointly modeling compositional outcomes and environmental mixtures using a Dirichlet distribution with a multinomial logit link function. The proposed approach, named Dirichlet-BWQS (DBWQS), allows for the simultaneous estimation of mixture weights associated with each exposure mixture component as well as the association between the overall exposure mixture index and each outcome proportion. We assess the performance of DBWQS regression on extensive simulated data and a real scenario where we investigate the associations between environmental chemical mixtures and DNA methylation-derived placental cell composition, using publicly available data (GSE75248). We also compare our findings with results considering environmental mixtures and each outcome component. Finally, we developed an R package \textit{xbwqs} where we made our proposed method publicly available (\textit{https://github.com/hasdk/xbwqs}). 
\end{abstract}

\section{Introduction}
Compositional data, also called fractional data, consist of vectors constrained onto the unit simplex (i.e., all vector components are non-negative proportions that sum to one). In health sciences, compositional data are encountered in various contexts, including nutrition and time allocation of daily activities \citep{Dumuid_2021}, relative abundance of microbial gut taxa \citep{lin_analysis_2020}, and cell type proportions of biological samples \citep{gao2019, saddiki_dna_2023}.
The naive and most widely used approach for compositional outcomes consists of fitting multiple `individual' regressions, one for each outcome proportion. Although most analyses used Poisson or negative binomial regressions to accommodate individual proportions, such methods do not take into account the simplical constraints that the outcome proportions sum to one, thus inducing bias in the estimated regression coefficients. This has led to the development of compositional regression models that aim to jointly model the relative proportions (i.e., multi-dimensional outcome) as a whole or a special \textit{mixture}. 

Since the outcome is considered a multi-dimensional variable, compositional regression models require the specification of a multi-dimensional distribution to jointly model the regression coefficients associated with each outcome proportion.  Methods for compositional regression can be broadly classified into two categories: (1) methods based on log-ratio transformations that map the compositional outcome to the Aitchison geometric space which allows for the application of multivariate linear regression analysis on the simplex \citep{aitchison_statistical_1986, Boogaart_Tolosana-Delgado_2013}; and (2) methods based on modeling the compositional response as a Dirichlet distribution via an appropriate link function such as the multinomial logit link \citep{Hijazi_Jernigan_2009, VanderMerwe_2019, ascari_multivariate_2023}. Methods based on log-ratio transformations provide regression coefficients with no straightforward interpretation as effect estimates since the regression model itself is fitted in the transformed space; while the second class of models provides regression coefficients that can readily be interpreted as changes in relative (or absolute) proportions for a unit increase in exposure. The major difference between the two approaches lies in the coefficient estimate bias, which is more pronounced from transformation-based approaches than those models with a multinomial logit link function in the presence of skewness and non-normality of the transformed outcomes. Although compositional regression models based on the Dirichlet distribution have been extensively studied both in frequentist \citep{Hijazi_Jernigan_2009, Ankam2019} and Bayesian settings \citep{DaSilva_Rodrigues_2015,VanderMerwe_2019,ascari_multivariate_2023}, no study has investigated the joint effect of multiple exposures when the outcome is compositional. 

In this work, we focus on compositional regression settings in which we investigate the joint effect of multiple exposures (namely, exposure mixture) on a compositional outcome of interest. 
The Bayesian weighted quantile sum (BWQS) approach is a supervised mixture approach that can evaluate the association between exposure mixtures and a single outcome of interest \citep{Colicino_2020}. This approach extends the frequentist weighed quantile sum (WQS) approach \citep{Carrico_2015, Czarnota_2015} to the Bayesian setting, overcoming some of its limitations.
We propose the Dirichlet-BWQS (DBWQS), an extension of BWQS regression to model the compositional outcome via a multinomial logit link. 

The manuscript is organized as follows: Section~\ref{sec:dir-reg} introduces notation and the Dirichlet regression model with its alternative representation and using the multinomial logit link; Section~\ref{sec:dbwqs-reg} builds on the Dirichlet model from Section~\ref{sec:dir-reg} to construct the Dirichlet-BWQS regression model; Section~\ref{sec:sim} describes the simulation experiments and reports the performance of our proposed approach; Section~\ref{sec:real-app} applies DBWQS to real data and analyzes the association between placental chemical mixtures and placental cell type composition; finally, Section~\ref{sec:disc} summarizes the simulation and real data results, and provides a discussion of limitations and potential future applications.

\section{Dirichlet Regression}\label{sec:dir-reg}

Consider a K-dimensional compositional outcome $\pmb{y} = (y_1, ..., y_K)$ consisting of proportions summing up to one. The Dirichlet distribution can be used to model $\pmb{y}$ with a real, positive, K-dimensional parameter vector $\pmb{\alpha} = (\alpha_1, ..., \alpha_K)$ as follows,
\begin{equation*}
    p(\pmb{y}) \sim Dirichlet(\pmb{\alpha}) = \frac{\Gamma(\sum_{k=1}^K \alpha_k)}{\prod_{k=1}^K \Gamma(\alpha_k)} \prod_{k=1}^K y_k^{\alpha_k-1},
\end{equation*}

where $\sum_{k=1}^K y_k=1$, $y_k > 0$, and $\alpha_k > 0$ for all $k=1,...,K$.

The alternative representation of the Dirichlet distribution aims at decoupling the $\pmb{\alpha}$ parameter vector such that the mean and precision of $\pmb{y}$ can be estimated separately; this formulation also allows for a more intuitive interpretation of the parameters in the context of association studies \citep{VanderMerwe_2019}.

The alternative representation of the Dirichlet distribution, parameterized by mean vector $\pmb{\mu} = (\mu_1, ..., \mu_k)$ and precision $\phi$, is
\begin{equation*}
    p(\pmb{y}) \sim Dirichlet(\pmb{\mu}, \phi) = \frac{\Gamma(\phi)}{\prod_{k=1}^K \Gamma(\phi . \mu_k)} \prod_{k=1}^K y_k^{(\phi . \mu_k)-1},
\end{equation*}
where $\phi = \sum_{k=1}^K \alpha_k > 0 , \ \mu_k = E[y_k] = \frac{\alpha_k}{\sum_{k=1}^K \alpha_k} = \frac{\alpha_k}{\phi} \implies \alpha_k = \phi . \mu_k, \ \sum_{k=1}^K \mu_k = 1$.

We now leverage this alternative Dirichlet representation and the Bayesian framework to describe compositional regressions including a set of independent predictors.

Consider a regression model where the outcome $\pmb{y}=(y_1,...,y_K)$ is a $K$-dimensional vector of proportions summing to one, and $\pmb{x}=(x_1, ..., x_J)$ is a $J$-dimensional vector of independent predictors. We now utilize the alternative representation of the Dirichlet distribution to perform a compositional regression with a \textit{multinomial-logit link}. We aim to estimate the mean vector $\pmb{\mu}=(\mu_1, ...,\mu_K)$ and the global precision parameter $\phi$. 

For each mean parameter $\mu_k$ ($k=1,...,K$), let $\pmb{\beta}_k$ be the $J$-dimensional vector of regression coefficients associated with the main set of independent predictors $\pmb{x}$, and let $\phi$ be the scalar global precision parameter.

For i.i.d observations $\{\pmb{y}_i, \pmb{x}_i\}_{(i=1,...,N)}$, the Dirichlet regression can be written with multinomial-logit link as:
\begin{align*}
    p(\pmb{y}_i | \pmb{x}_i) &\sim Dirichlet(\pmb{\mu}_i, \phi), \\
    \text{where} \ \mu_{i,k} &= \frac{\exp(\pmb{x}_i^T \pmb{\beta}_{k})}{\sum_{c=1}^K \exp(\pmb{x}_i^T \pmb{\beta}_{c})} , \ \text{for} \ k=1,...,K.
\end{align*}

We defined $(k=1)$ to be a \textit{base} or \textit{reference} category whereby the vector of regression coefficients corresponding to this outcome category are set to zero ($\pmb{\beta}_1 = \pmb{0}$). This transformation eliminates the redundancy stemming from the simplical constraints of the compositional outcomes, while providing an approach to estimate coefficients that can be interpreted as changes in relative or absolute proportions (more details regarding the interpretation of effect estimates are presented in Section~\ref{sec:dbwqs-interpret-effects}). In here, we adopted a Bayesian specification for the Dirichlet regression model where we defined Normal priors on the regression coefficients $\beta_{k,j}$ and Gamma priors on the precision parameter $\phi$.

\section{Dirichlet-BWQS (DBWQS) Regression}\label{sec:dbwqs-reg}
% \paragraph{Dirichlet-BWQS (DBWQS) Regression}
We now build upon the Dirichlet regression model to extend Bayesian Weighted Quantile Sum (BWQS) regression to compositional outcome settings (i.e., DBWQS).

Consider a set of $M$ co-occurring, potentially correlated exposures $\pmb{z}=(z_1, ..., z_M)$ individually ranked in quantiles denoted as $F{^{-1}}{_{Z_{m}}}(z_{i,m})=q_{i,m}$ for the $i$-th individual's $m$-th exposure level with ($m=1,...,M$) and ($i=1,...,N$). Let $\boldsymbol{w}$=($w_1,...,w_M$) be the mixture weight vector lying on the Unit Simplex, with individual weight being non-negative ($w_m \geq 0, \forall m$) and summing to 1 $(\sum_{m=1}^{M} w_m = 1)$, $\boldsymbol{\theta}$=($\theta_1, ..., \theta_K$) the regression coefficient vector associated with the BWQS mixture for the mean parameters, and $\phi$ the global precision parameter. The BWQS mixture index for a given individual is $S_i=\sum_{m=1}^M q_{i,m} w_m$.

For i.i.d observations $\{\pmb{y}_i, \pmb{x}_i, \pmb{q}_i\}_{i=1,...,N}$, the DBWQS model can be written as
\begin{align*}
    p(\pmb{y}_i | \pmb{x}_i, \pmb{q}_i) &\sim Dirichlet(\pmb{\mu}_i, \phi), \\
    \mu_{i,k} &= \frac{\exp(S_i \theta_k  + \pmb{x}_i^T \pmb{\beta}_{k})}{\sum_{c=1}^K \exp(S_i\theta_c + \pmb{x}_i^T \pmb{\beta}_{c})} , \ \text{for} \ k=1,...,K,\\
    S_i &= \sum_{m=1}^M q_{i,m} w_m,\\
    \pmb{\beta}_1 &= \pmb{0} \ \textit{and} \ \theta_1 = 0 \quad \ \textit{(base or reference category)},\\
    \pmb{w} &\sim Dirichlet(\pi_1, ..., \pi_M), \\
    \pi_m &\sim Gamma(a_{\pi},b_{\pi}), \quad \text{for} \ m=1,...,M,\\
    \beta_{k,j} &\sim Normal(0, \sigma_{\beta}), \quad \text{for} \ k=2,...,K, and \ j=1,...,J, \\
    \theta_k &\sim Normal(0, \sigma_{\theta}), \quad \text{for} \ k=2, ..., K, \\
    \phi &\sim Gamma(a_{\phi},b_{\phi}), \\
    &a_{\pi}>0, \ b_{\pi}>0, \ a_{\phi}>0, \ b_{\phi}>0.
\end{align*}

\paragraph{Prior probability distributions and hyperparameters}

We specified weakly informative Normal prior probability distributions  centered at zero with large values for the standard deviation (e.g., $\sigma_{\beta} = 100$) on all regression coefficients ($\pmb{\theta}$ and $\pmb{\beta}$) except those corresponding to the reference category ($k=1$), which are set to zero. Following the notation of prior studies ~\citep{ascari_multivariate_2023, Colicino_2020}, we included a Gamma prior distribution on the precision parameter $\phi$ with small hyperparameter values ($a_{\phi}=b_{\phi}=0.001$) to induce a large variability; and Dirichlet prior probability distributions with parameters ($\pi_1, ..., \pi_M$) for the mixture weights ($w_1, ..., w_M$). The Dirichlet prior was employed because it is conjugate, naturally satisfies the $M$-dimensional unit simplex (i.e., mixture weights that are non-negative and sum to one) and it simplifies the interpretation of findings. Additionally, we endow each mixture weight parameter ($\pi_m$, for $m=1,...,M$) with a Gamma prior probability distribution and moderate hyperparameter values ($a_{\pi}=a_{\pi}=2$); this allows for more diffuse mixture weights so that instances where some mixture components have very low or very high relative contributions (i.e., values close to 0 or 1) can be handled by the model.

\paragraph{Bayesian Inference in Stan}

Exact inference on the posterior of the DBWQS model (derived in supplementary materials) is intractable to compute. Therefore, we used Markov Chain Monte Carlo algorithms to sample from the joint posterior distribution of interest. In particular, we implement our inference procedure for DBWQS in Stan with the No-U-Turn Sampler (NUTS), a variant of the Hamiltonian Monte Carlo (HMC) algorithm that allows for adaptive tuning of the sampling parameters (e.g., number steps and step size) at each iteration \citep{hoffman2014}. Specifically, the step size is tuned during warm-up and then held fixed during sampling. In all experiments, we ran $2$ Monte Carlo chains, each using $10,000$ iterations with $2,000$ employed for the warm-up phase; and we reported the effective sample size and $\hat{R}$ along with autocorrelation plots to diagnose convergence. We developed an R package \textit{xbwqs} where the DBWQS method is made publicly available (\textit{https://github.com/hasdk/xbwqs}).

\paragraph{Interpretation of Effect Estimates from DBWQS Model}\label{sec:dbwqs-interpret-effects}

The multinomial-logit link used in the definition of the DBWQS model allows for a straightforward interpretation of the regression coefficients $\theta_k$ (for $k=2,...,K$) associated with the weighted mixture index in terms of changes in \textit{relative proportions} of a given outcome category with respect to the reference category ($k=1$). However, in some applications, it is of interest to estimate the change in absolute proportions of each outcome proportion for a unit increase in exposure. In this section, we show how the change in absolute proportions can be represented as a function of the regression coefficients $\pmb{\theta}$.

Recalling that $S$ is the weighted mixture index and $\pmb{x}$ the adjustment covariates, we have the following expression for $\mu_k$, the mean absolute proportion of each outcome category (k=1,...,K),
\begin{align}
    \text{For} \ (k=1), \quad \mu_{1} &= \frac{\exp(S . 0 + \pmb{x}^T . 0)}{1 + \sum_{c=2}^K \exp(S .\theta_c + \pmb{x}^T \pmb{\beta}_{c})} = \frac{1}{1 + \sum_{c=2}^K \exp(S. \theta_c +\pmb{x}^T \pmb{\beta}_{c})}, \\
    \text{For} \ (k=2,...,K), \quad \mu_{k} &= \frac{\exp(S. \theta_k  + \pmb{x}^T \pmb{\beta}_{k})}{1 + \sum_{c=2}^K \exp(S. \theta_c +\pmb{x}^T \pmb{\beta}_{c})}. 
\end{align}

Applying the natural logarithm on both sides of each equation, we get
\begin{align}
    \text{For} \ (k=1), \quad \log (\mu_{1}) &= 0 - \log\left(1 + \sum_{c=2}^K \exp(S. \theta_c +\pmb{x}^T \pmb{\beta}_{c})\right), \\
    \text{For} \ (k=2,...,K), \quad \log( \mu_{k}) &= S .\theta_k  + \pmb{x}^T \pmb{\beta}_{k}  - \log\left( 1+ \sum_{c=2}^K \exp(S.\theta_c + \pmb{x}^T \pmb{\beta}_{c}) \right).
\end{align}

Let $\mu^*_k$ denote the estimated absolute proportion given a one unit increase in exposure S (i.e., we replace S by (S+1) in the equations above) and holding the covariates fixed, we get,
\begin{align}
    \text{For} \ (k=1), \quad \log (\mu^*_{1}) &= 0 - \log\left(1 + \sum_{c=2}^K \exp(S. \theta_c + \theta_c +\pmb{x}^T \pmb{\beta}_{c})\right), \\
    \text{For} \ (k=2,...,K), \quad \log( \mu^*_{k}) &= S .\theta_k +\theta_k  + \pmb{x}^T \pmb{\beta}_{k}  - \log\left(1+ \sum_{c=2}^K \exp(S.\theta_c + \theta_c + \pmb{x}^T \pmb{\beta}_{c}) \right).
\end{align}

Next, taking the difference between $\log(\mu^*_k)$ and $\log(\mu_k)$, we get
\begin{align}
    \text{For} \ (k=1), \quad \log (\mu^*_{1}) - \log(\mu_1) &= \log \left( \frac{1 + \sum_{c=2}^K \exp(S. \theta_c +\pmb{x}^T \pmb{\beta}_{c})}{1 + \sum_{c=2}^K \exp(S. \theta_c +\pmb{x}^T \pmb{\beta}_{c}) . \exp(\theta_c)} \right) \triangleq \log (b_{ref}) \nonumber \\
    \implies \frac{\mu^*_1}{\mu_1} &= b_{ref}  \\
    \text{For} \ (k=2,...,K), \quad \log (\mu^*_{k}) - \log(\mu_k) &= \theta_k + \log (b_{ref}) \nonumber \\
    \implies \frac{\mu^*_k}{\mu_k} &= \exp(\theta_k) \ . \ b_{ref} 
\end{align}

The last two equations allow us to quantify the change in absolute proportions of each cell type associated with a one unit increase in mixture index S, averaged over all participants in the data set. Specifically, if $\frac{\mu^*_k}{\mu_k}$ is equal to 1, then we conclude that there is no change in the absolute proportions of category $k$ for a unit increase in exposure $S$. If it is strictly different than 1, then an increase in exposure is associated with a change of $(\frac{\mu^*_k}{\mu_k} - 1) \times 100\%$ in absolute proportions of outcome category $k$. In practice, we can calculate the average $b_{ref}$ and $(\exp(\theta_k) \times b_{ref})$ from posterior samples and get credible intervals for the effect of interest in absolute proportion scale.

\section{Simulation Studies}\label{sec:sim}

We assess the performance of the DBWQS regression on simulated data scenarios with varying parameter settings. Each setting is replicated 100 times to obtain Monte Carlo estimates of the mean, root mean squared error (RMSE), mean absolute error (MAE), mean standard errors, and coverage probabilities of the credible sets. In the next section, we also compared results from the DBWQS regression with those obtained from the compatible BWQS model, which accommodates the exposure mixture but does not take into consideration the compositional outcome. 
All models are run using the \textit{dbwqs} package, which depends on the \textit{rstan} package (v2.32.6; \cite{Stan_Development_Team2020-sz}) in R programming language \citep{Rlang}.

\paragraph{Data generating model} Following the same notation and indices as before, the data generating model for our simulated scenarios is described as follows,

\begin{enumerate}
    \item Set true regression coefficients $\pmb{\theta} = (\theta_1, ..., \theta_K)$ and mixture weights $\pmb{w} = (w_1, ..., w_M)$.
    \item Simulate $J$ covariates $\pmb{x}$ by drawing $N$ samples independently from Normal$(0,1)$, and their associated regression coefficients $\pmb{\beta}_k$ (for $k=2,...,K$) independently from Uniform$(-1,1)$.
    \item Simulate $M$ exposures by drawing $N$ samples from a multivariate normal centered at \pmb{0}, with variance $=1$, and covariance $=\rho$; calculating the quartiles of each exposure yields $\pmb{q}$. 
    \item Calculate the weighted quartile sum $S_i = \sum_{m=1}^M q_{i,m} w_m$ for ($i=1,...,N$), where ($w_1,...,w_m$) are the pre-specified mixture weights.
    \item Using the simulated quantities from previous steps, calculate the mean vectors $\mu_{i,k}$ for $k=1,...,K$ and $i=1,...,N$ using the Dirichlet-BWQS regression model equations (17)-(18) from Section~\ref{sec:dbwqs-reg}.
    \item Using a pre-specified value for the global precision parameter $\phi$ and the mean vector $\mu_{i,k}$, calculate the Dirichlet parameter vectors as $\alpha_{i,k} = \phi . \mu_{i,k}$ for $k=1,...,K$, and $i=1,...,N$.
    \item Finally, simulate the compositional outcome $\pmb{y}_i$ by drawing from the Dirichlet$(\alpha_{i,k})$ distribution.
\end{enumerate}

\paragraph{Simulation Parameters} We summarize the parameter settings for the different simulation scenarios in Table~\ref{table:1}. The chosen values of the simulation parameters are based on realistic scenarios often encountered in epidemiologic studies involving mixtures of chemical exposures \cite{cowell2021, saddiki_dna_2023, indiaaldana2024}.

\begin{table}[h!]
\centering
\begin{tabular}{ |c|c| } 
 \hline
 \textbf{Parameter} & \textbf{Values}  \\
 \hline 
  Sample size ($n$) & $[150, 300]$  \\ 
  Outcome categories ($K$) & $[3, 6, 9]$ \\ 
  Number of exposures ($M$) & $[3, 6, 9]$ \\
  Number of covariates ($J$) & $[0, 3]$ \\
  Global precision ($\phi$) & $5$ \\
  Exposure correlation ($\rho$) & $[0.3, 0.6]$ \\
  Monte Carlo repetitions & 100 \\
 \hline
\end{tabular}
\caption{Table summarizing parameter values considered for simulation experiments.}
\label{table:1}
\end{table}

\paragraph{True coefficients and weights} We set fixed values for the regression coefficients $\pmb{\theta}$ and mixture weights $\pmb{w}$ for each setting of $K$ and $M$, respectively. These values are summarized in Table~\ref{tbl:true_simparams}.

\begin{table}[!ht]
\centering
\begin{tabular}{ |c|c|c| } 
 \hline
 Regression coefficients ($\pmb{\theta}$) & $K=3$ & $\pmb{\theta} = [0, 0, -0.9]$ \\
 & $K=6$ & $\pmb{\theta} = [0,0,-0.9,-0.5,0.8,0.9]$ \\
 & $K=9$ & $\pmb{\theta} = [0,0, -0.5, -0.85, 0.8, 0.9, -0.2, 0.1, -0.3]$\\
 \hline
 Mixture weights ($\pmb{w}$) & $M=3$& $\pmb{w} = [0.8,0,0.2]$\\
 & $M=6$ & $\pmb{w} = [0.3,0.1,0.1,0.1,0.2,0.2]$ \\
 & $M=9$ & $\pmb{w} = [0.4,0.3,0.2,0.1,0,0,0,0,0]$ \\
 \hline
\end{tabular}
\caption{Table summarizing true simulated values for the regression coefficients $\pmb{\theta}$ and mixture weights $\pmb{w}$ for each setting of $K$ and $M$.}
\label{tbl:true_simparams}
\end{table}

\subsection{Simulation Results}

\begin{figure}[!htb]
\centering
\includegraphics[width=\textwidth]{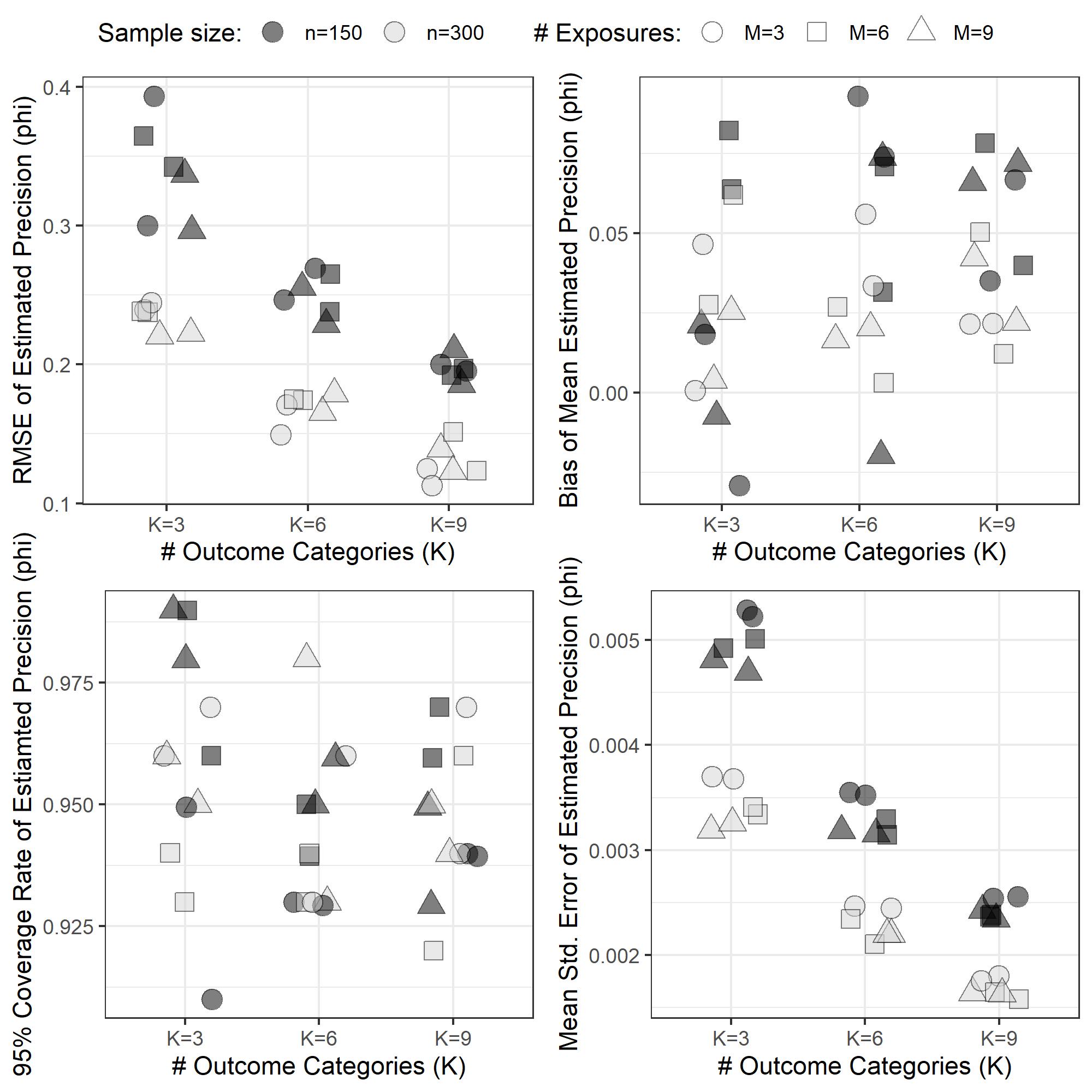}
\caption{Summary of simulation performance results for the estimated global precision $\phi$ across different sample sizes ($n$), number of exposures ($M$), number of outcomes ($K$), and exposure correlations ($\rho$). The number of covariates are fixed at $J=0$. All performance estimates shown are averaged across 100 Monte Carlo repetitions.} 
\label{fig:simres-phi-J0}
\end{figure}

\begin{figure}[!htb]
\centering
\includegraphics[width=\textwidth]{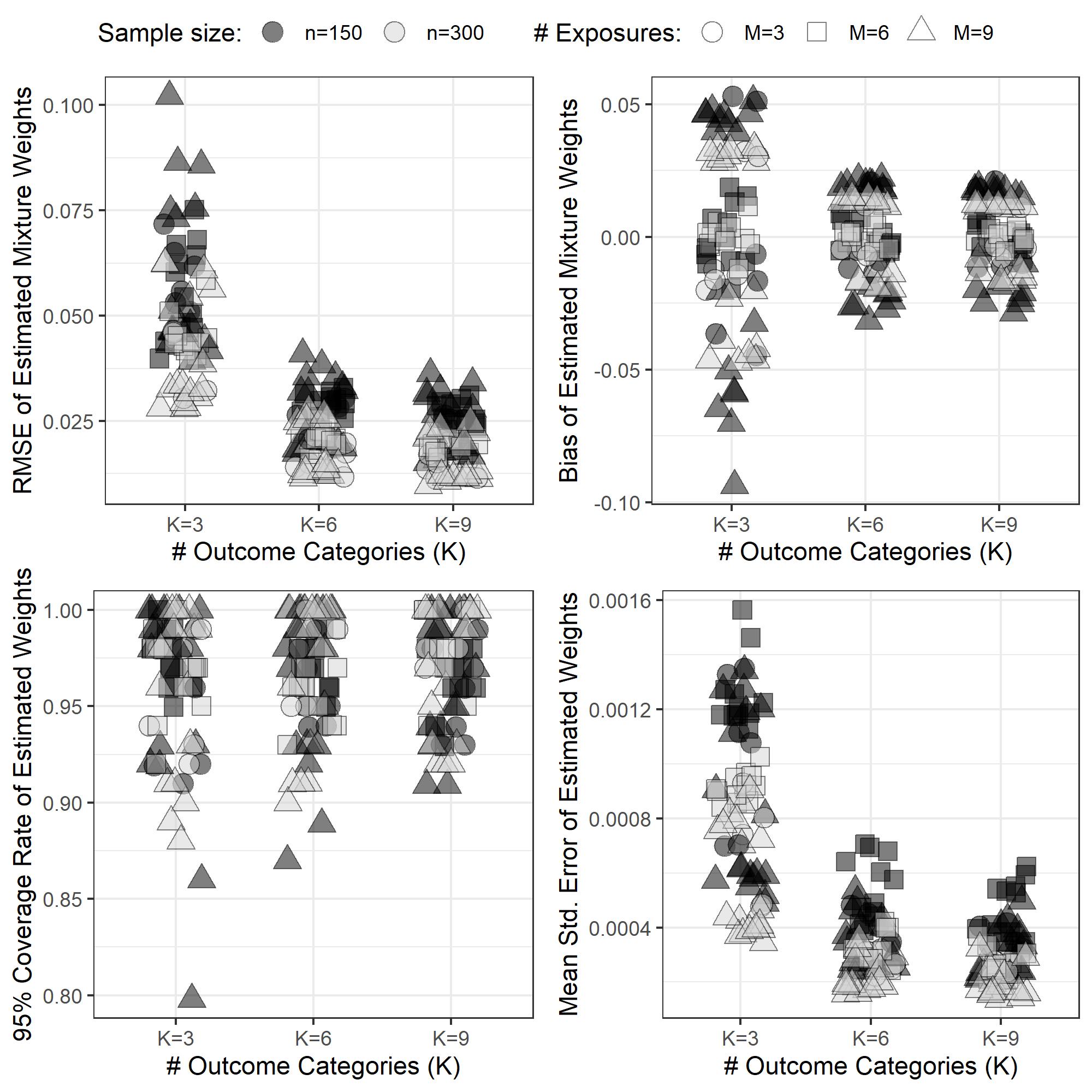}
\caption{Summary of simulation performance results for the estimated mixture weights across different sample sizes ($n$), number of exposures ($M$), number of outcomes ($K$), and exposure correlations ($\rho$). The number of covariates are fixed at $J=0$. All performance measures shown are averaged across 100 Monte Carlo repetitions.} 
\label{fig:simres-wt-J0}
\end{figure}

\begin{figure}[!htb]
\centering
\includegraphics[width=\textwidth]{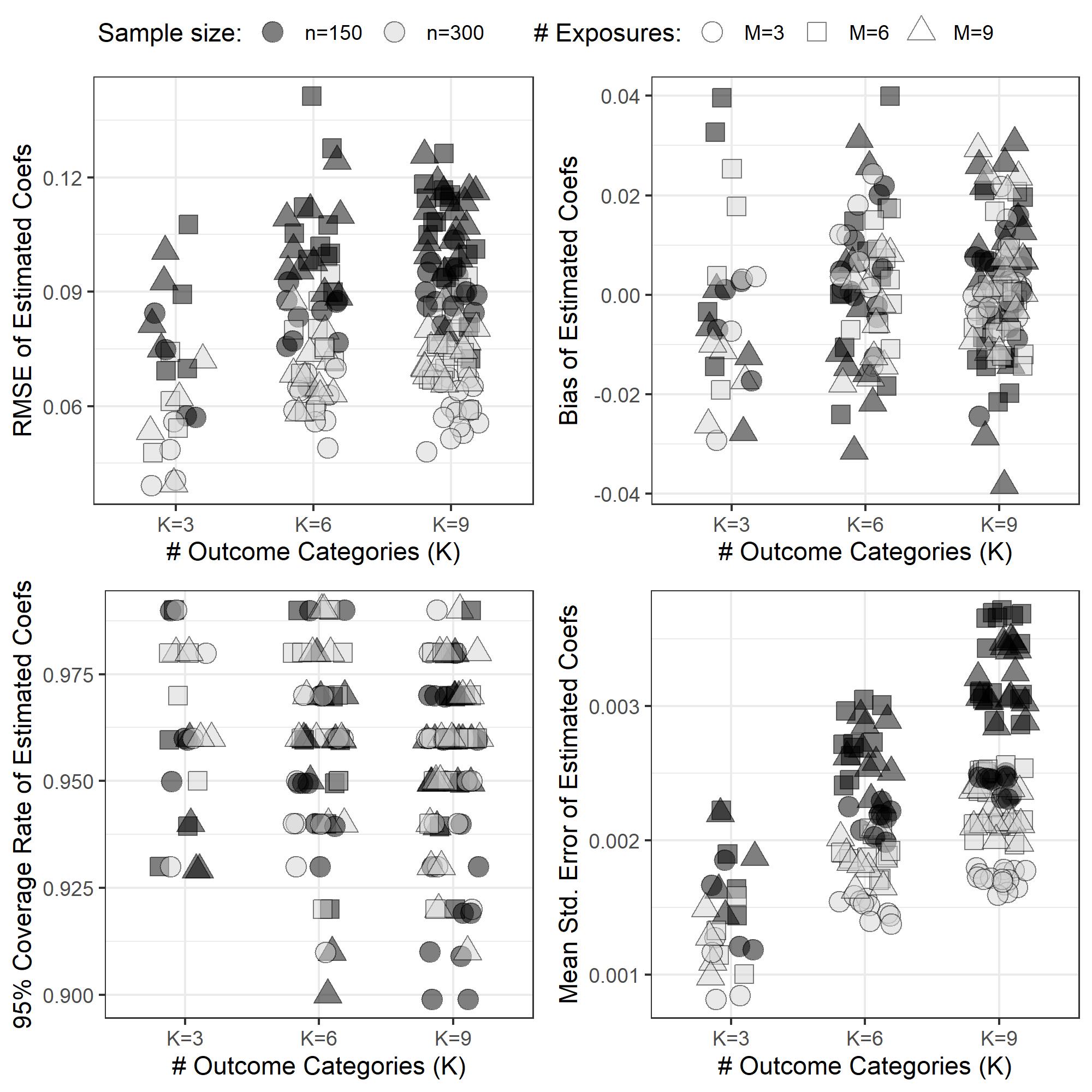}
\caption{Summary of simulation performance results for the estimated regression coefficients across different sample sizes ($n$), number of exposures ($M$), number of outcomes ($K$), and exposure correlations ($\rho$). The number of covariates are fixed at $J=0$. All performance estimates shown are averaged across 100 Monte Carlo repetitions.} 
\label{fig:simres-coefs-J0}
\end{figure}

Findings from simulated scenarios showed that both the strength of the exposure correlation $\rho$ and the number of independent covariates $J$ minimally impacted the performance of DBWQS regression. For example, the average RMSE and bias for the regression coefficients $\pmb{\theta}$ were $0.087$ and $0.001$ across all scenarios with $\rho=0.3$ compared to $0.080$ and $0.002$, respectively, for those with $\rho=0.6$. Similarly, the average RMSE and bias for the regression coefficients $\pmb{\theta}$ were $0.082$ and $0.001$ across all scenarios with no covariates ($J=0$) compared to $0.084$ and $0.003$, respectively, for those with 3 covariates ($J=3$). The standard errors remained small across different simulation scenarios, indicating high precision in the parameter estimates even at low to moderate sample sizes and higher model complexity.

We now focus on how the sample size ($n$), number of exposures ($M$) and number of outcomes ($K$) affect DBWQS model performance, assuming a fixed setting for $\rho$ and $J$ across scenarios. All reported performance results are averaged across 100 Monte Carlo repetitions. 

Overall, the model estimates for the regression coefficients $\pmb{\theta}$, mixture weights $\pmb{w}$ and global precision $\phi$ achieve low Root Mean Square Error (RMSE) and proper coverage rates across all simulated scenarios. As expected, scenarios with larger sample size ($n=300$) achieve slightly lower RMSE and better coverage rates than their smaller sample size counterparts ($n=150$).

We observe that the model performance improves (especially in terms of RMSE and bias) for higher number of outcome categories and larger sample size (\textbf{\Cref{fig:simres-phi-J0,fig:simres-wt-J0,fig:simres-coefs-J0}}). Finally, for low to moderate number of exposures (i.e., $3$-$6$ exposures), the model performance remains consistent across different number of outcome categories (i.e., $3$-$9$), especially in terms of the bias and RMSE of the estimated coefficients (\textbf{\Cref{fig:simres-phi-J0,fig:simres-wt-J0,fig:simres-coefs-J0}}).

\subsection{Competing method: BWQS regression on single outcome proportions}

DBWQS achieves adequate performance in estimating the true effect estimates of the exposure mixtures on the compositional outcome in the simulated data. An alternative environmental mixture approach would be to model the proportion of each outcome component as a univariate outcome in a separate Beta regression model. We compare DBWQS, with the BWQS regression that handles a single outcome proportion with exposure mixture. 

Let $p$ be the proportion of a single outcome, we consider the 2-dimensional composition $[p, 1-p]$ as our new outcome and perform DBWQS regression; in this special case of the compositional outcome being 2-dimensional, the Dirichlet distribution collapses to a Beta distribution.

In this section, we refer to using DBWQS to jointly model the compositional outcome as the \textit{joint outcome} approach, and using multiple BWQS models (one for each component of the composition separately) as the \textit{individual outcome} approach. Using the simulated data scenarios previously described, we employ BWQS using the individual outcome approach and compare the performance with the results obtained from the joint outcome DBWQS approach. Performance is reported in terms of RMSE, bias, and coverage rates in estimating the true regression coefficients associated with the simulated exposure mixtures.

\begin{figure}[!htb]
\centering
\includegraphics[width=\textwidth]{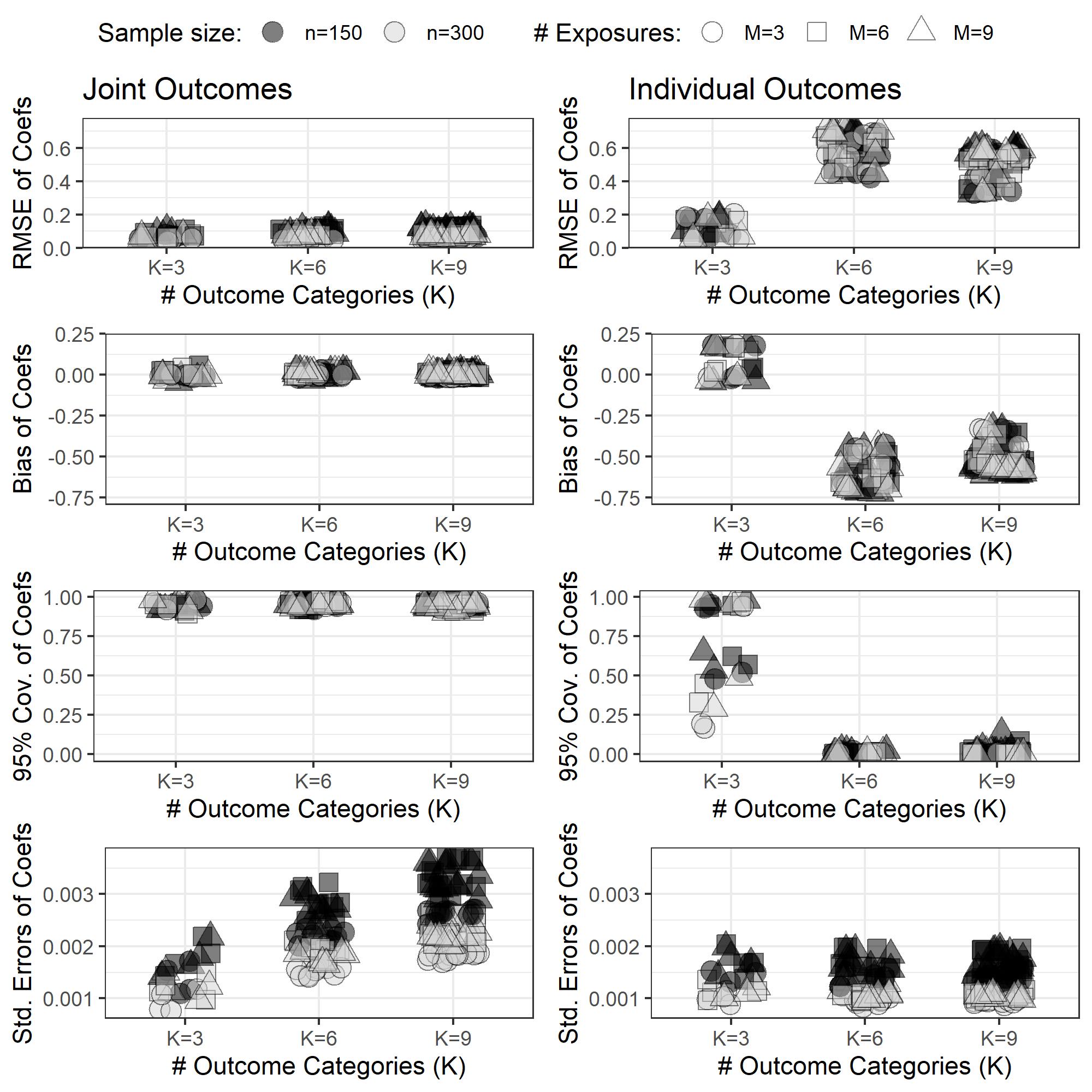}
\caption{Comparison of simulation performance results between DBWQS using \textit{joint} versus \textit{individual} outcome approaches. The performance was reported for the estimated regression coefficients in relative proportion scale across different sample sizes ($n$), number of exposures ($M$), number of outcomes ($K$), and exposure correlations ($\rho$). The number of covariates are fixed at $J=3$. All performance estimates shown are averaged across 100 Monte Carlo repetitions.} 
\label{fig:simres-compare-J3}
\end{figure}

Results in Figure~\ref{fig:simres-compare-J3} show a clear and consistent pattern: the performance of BWQS regressions using the individual outcomes approach degrades significantly (increased RMSE, bias deviates away from zero, and coverage drops to zero) as the number of outcome categories increases from $K=3$ to $K=9$, while the performance of the DBWQS model remains consistently adequate across the different simulation settings. The standard errors were low and well contained across the different simulation scenarios indicating high precision in parameter estimates for both approaches. The results of this comparison provide further evidence supporting the importance of jointly modeling compositional outcomes, as it highlights the potential biases and false conclusions that could be incurred from failing to account for the dependency structure between components of a compositional outcome.

\section{Real case scenario: co-occuring chemical exposures and placental cell type composition}\label{sec:real-app}

We leveraged data of 141 mother-infant pairs from the Rhode Island Child Health Study (RICHS) cohort with both placental chemical exposures and large-scale placental DNA methylation levels, publicly available through the Gene Expression Omnibus (GEO) repository (GSE75248; \cite{paquette2016}). Using the DBWQS regression, we evaluated the association between a mixture of co-occuring correlated chemicals assessed in placenta tissues and DNA methylation-derived placenta cell type composition, while adjusting for confounding variables.

\textbf{Study Population.} RICHS recruited mother-infant pairs from the Women and Infants Hospital of Rhode Island (2010-2013). Newborns considered LGA (large for gestational age) and SGA (small for gestational age) were matched to AGA infants (adequate for gestational age) on sex, gestational age ($\pm 3$ days), and maternal age ($2$ years). Exclusion criteria included:  maternal age $<18$ years, maternal life-threatening medical complication, twin pregnancies, infant congenital or chromosomal abnormality. Medical information was obtained from a structured chart review, and demographic, lifestyle, and exposure histories were collected from interviewer-based questionnaires. 

\textbf{Placenta tissue collection and DNA extraction}. RICHS placental tissue (free of maternal decidua) was collected within 2 hours of birth, three fragments from each of four placental quadrants were placed on RNAlater. After 72 hours, tissues were snap-frozen in liquid nitrogen, homogenized and stored $-80 \degree$C. DNA was extracted using the DNAeasy blood and tissue kit (QIAGEN, Inc, Valencia, CA) and bisulfite converted using the EZ DNA Methylation kit (Zymo, Irvine, CA).

\textbf{DNA methylation-derived placenta cell proportions}.  DNA methylation was profiled with the Illumina Infinium HumanMethylation450 (450K) BeadChip array, which includes approximately 450,000 genomic sites (i.e., CpG). We utilized the \textit{GEOquery R-package} \citep{davis2007} to obtain the normalized site-specific DNA methylation levels and phenotype data of RICHS participants. We then estimated the individual proportions of six cell types of placental tissues (syncytiotrophoblasts, trophoblasts, stromal, endothelial, Hofbauer and nucleated red blood cells (nRBCs))
using the reference-based cell type deconvolution method via the \textit{Planet R-package} \citep{yuan2021} and the constrained projection approach \citep{houseman2012}. 

\textbf{Chemical levels in placenta tissue.} Placental chemical concentrations were measured using standardized inductively coupled plasma mass spectrometry (ICP-MS) protocols at the Dartmouth Trace Elements Analysis laboratory \citep{tung2022}. The current study leveraged chemical concentrations (ng/g) for aluminum (Al), arsenic (As), chromium (Cr), cobalt (Co), lead (Pb), nickel (Ni), and selenium (Se), as previously described \citep{paquette2016, litzky2017, everson2018}. 

\begin{figure}[H]
\centering
\includegraphics[width=0.8\textwidth]{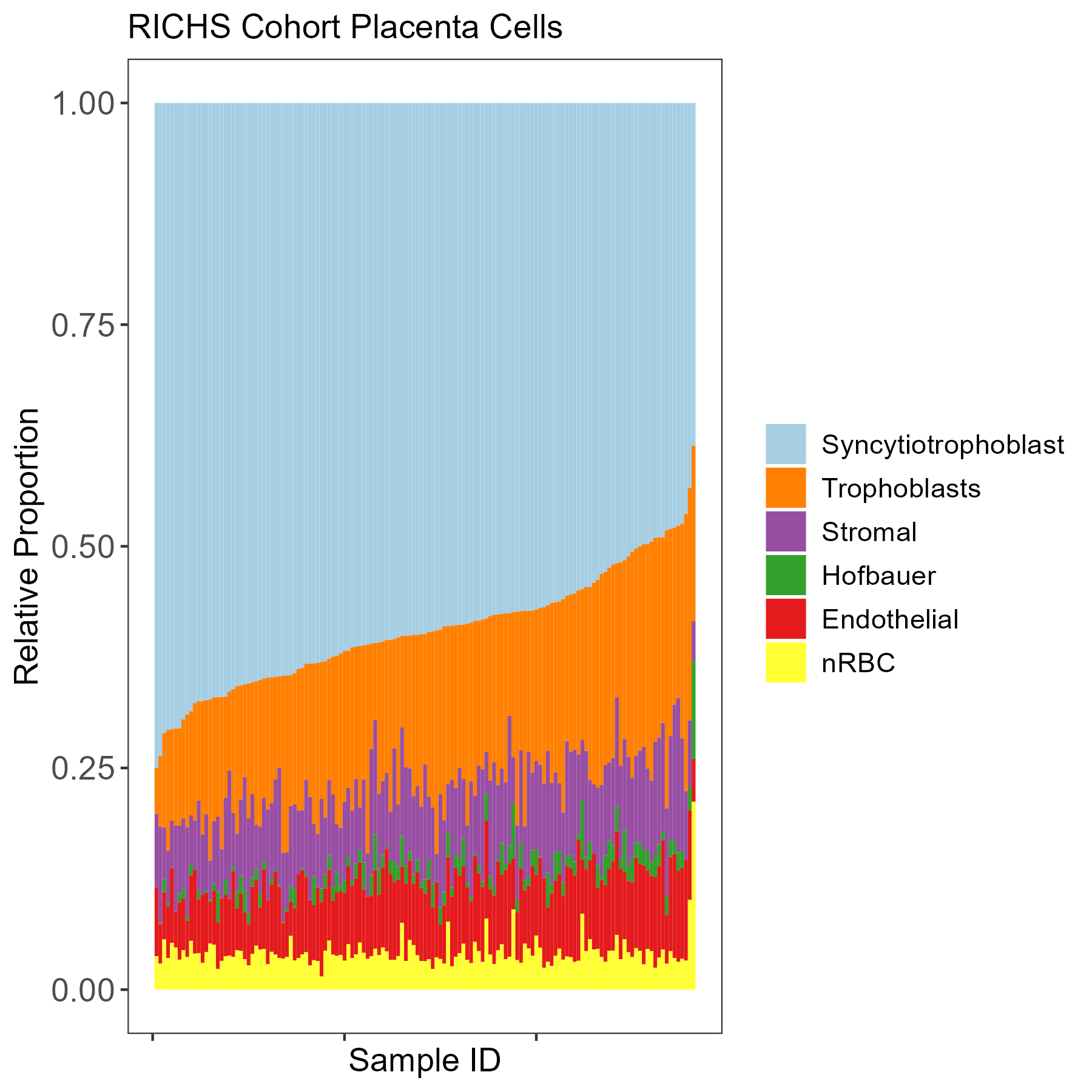}
\caption{DNA-methylation-derived placenta cell type composition for $n=141$ RICHS mother-child pairs. Individual samples are sorted by decreasing proportion of syncytiotrophoblasts, the most abundance proportion on average.} 
\label{fig-placenta-cellcomp-dist}
\end{figure}

\begin{figure}[H]
\centering
\includegraphics[width=0.8\textwidth]{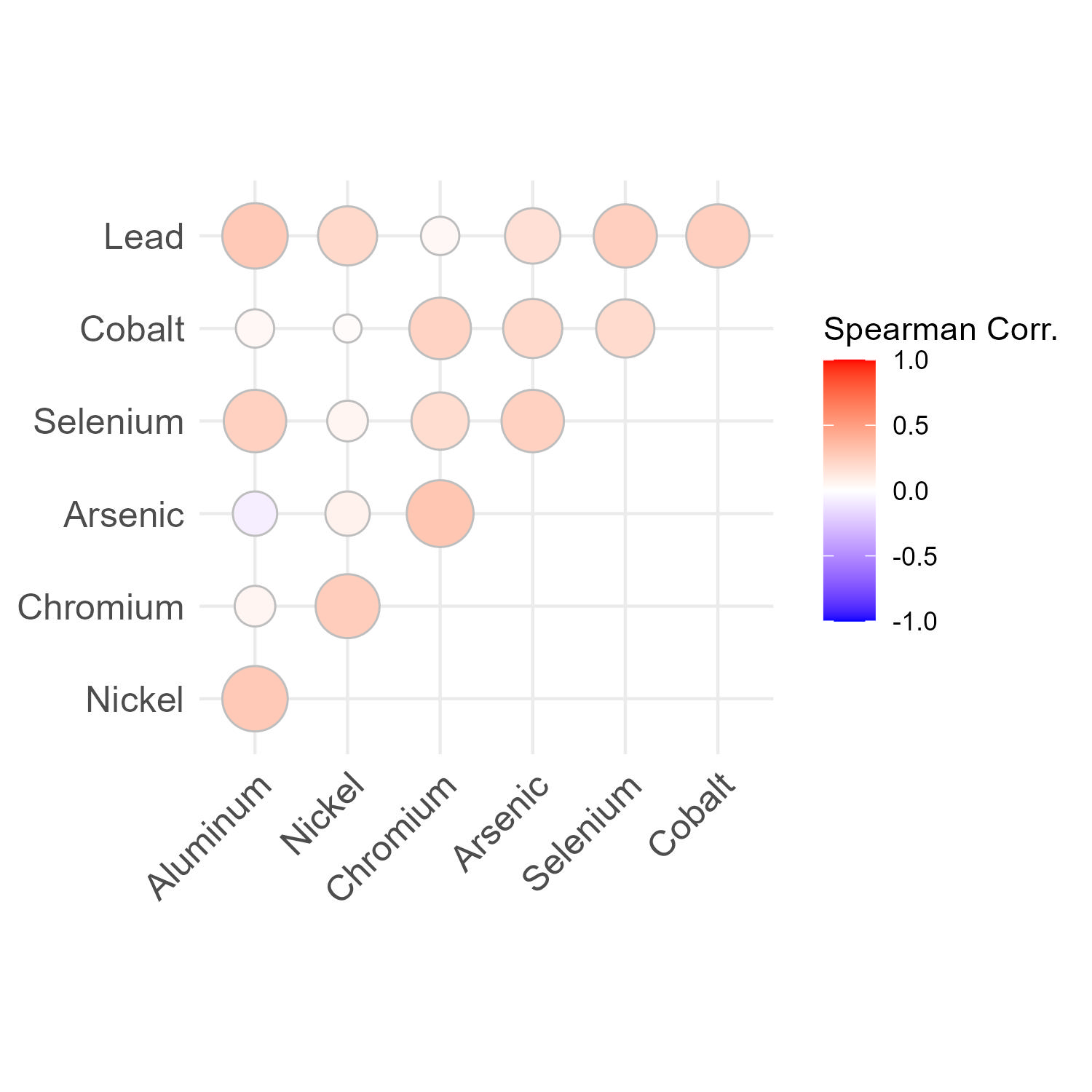}
\caption{Plot visualizing the Spearman correlation between the $\log_2$ concentrations of placenta chemicals in the RICHS cohort. The data shows positive correlation between a majority of the chemicals considered as exposures.} 
\label{fig-exposure-corr}
\end{figure}

\begin{table}[h!]
\centering
\begin{tabular}{ |c|c| } 
 \hline
   \textbf{Characteristic} & \textbf{Mean (SD) or N (\%)}   \\
   \hline
       \hline
       \multicolumn{2}{|c|}{\textbf{Mothers}}\\
  \hline 
   BMI (Kg/$m^2$) & 27 (6.8)  \\
   Age (Years) & 30.1 (5.4)  \\
   Smoking (Yes) & 22 (15.6\%)  \\
   Race (non-Hispanic, non-Black) & 104 (73.7\%)   \\
   Education (HS or less) & 34 (24.1\%)  \\
   Gestational diabetes (Yes) & 20 (14.2\%)  \\
   Fetal sex (Female) & 68 (48.2\%)  \\
  \hline 
  \multicolumn{2}{|c|}{\textbf{Placenta Chemicals} (ng/g, $log_2$ concentration)} \\
  \hline
  Arsenic (As) & -11.6 (7.9) \\
  Selenium (Se) & 8.1 (0.2) \\
  Chromium (Cr) & 4.3 (1) \\
  Cobalt (Co) & 1.8 (0.6) \\
  Lead (Pb) & 1.3 (1)  \\
  Aluminum (Al) & 6.2 (6.1)\\
  Nickel (Ni) & 1.3 (3.2)\\
  \hline
\end{tabular}
\caption{RICHS cohort characteristics ($n=141$).}
\label{tbl1-richs}
\end{table}

\subsection{Results}

Women were on average 30 years of age, and the majority of them were non-Hispanic non-Black ($73.7\%$), had a at least high-school education ($75.9\%$), were non-smokers ($84.4\%$), and were not diagnosed with gestational diabetes ($85.8\%$). Exposure levels assessed in placenta were log$_2$-transformed and reported in \textbf{Table~\ref{tbl1-richs}}. The Spearman correlations between the placenta chemicals considered in our analysis are shown in \textbf{Figure~\ref{fig-exposure-corr}}. \textbf{Figure~\ref{fig-placenta-cellcomp-dist}} shows the distribution of estimated placenta cell type compositions for each mother-child pair in the data set. 

\begin{figure}[!htb]
\centering
\includegraphics[width=0.6\textwidth]{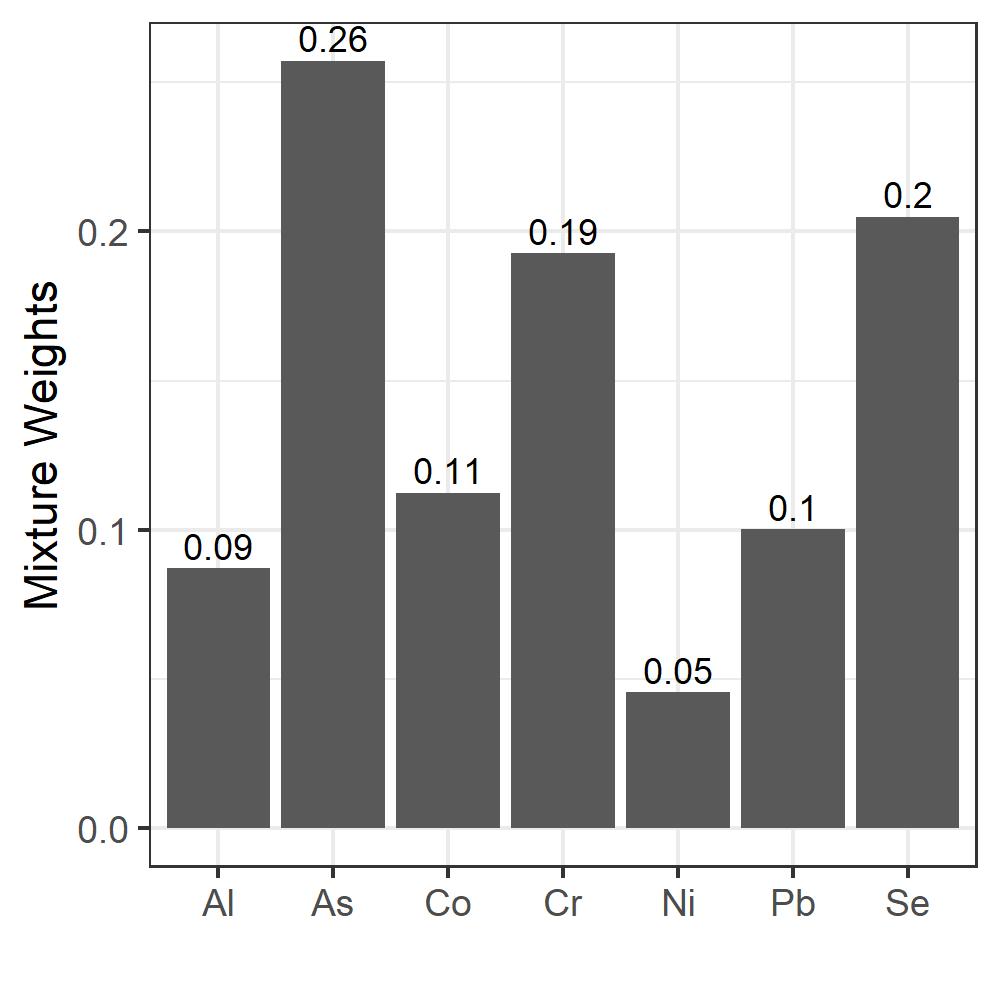}
\caption{Bar plot of DBWQS mean estimates of mixture weights associated with each chemical in the mixture: Aluminum (Al), arsenic (As), cobalt (Co), chromium (Cr), nickel (Ni), lead (Pb), and selenium (Se).} 
\label{fig:barplot-mix-weights}
\end{figure}

\begin{figure}[!htb]
\centering
\includegraphics[width=\textwidth]{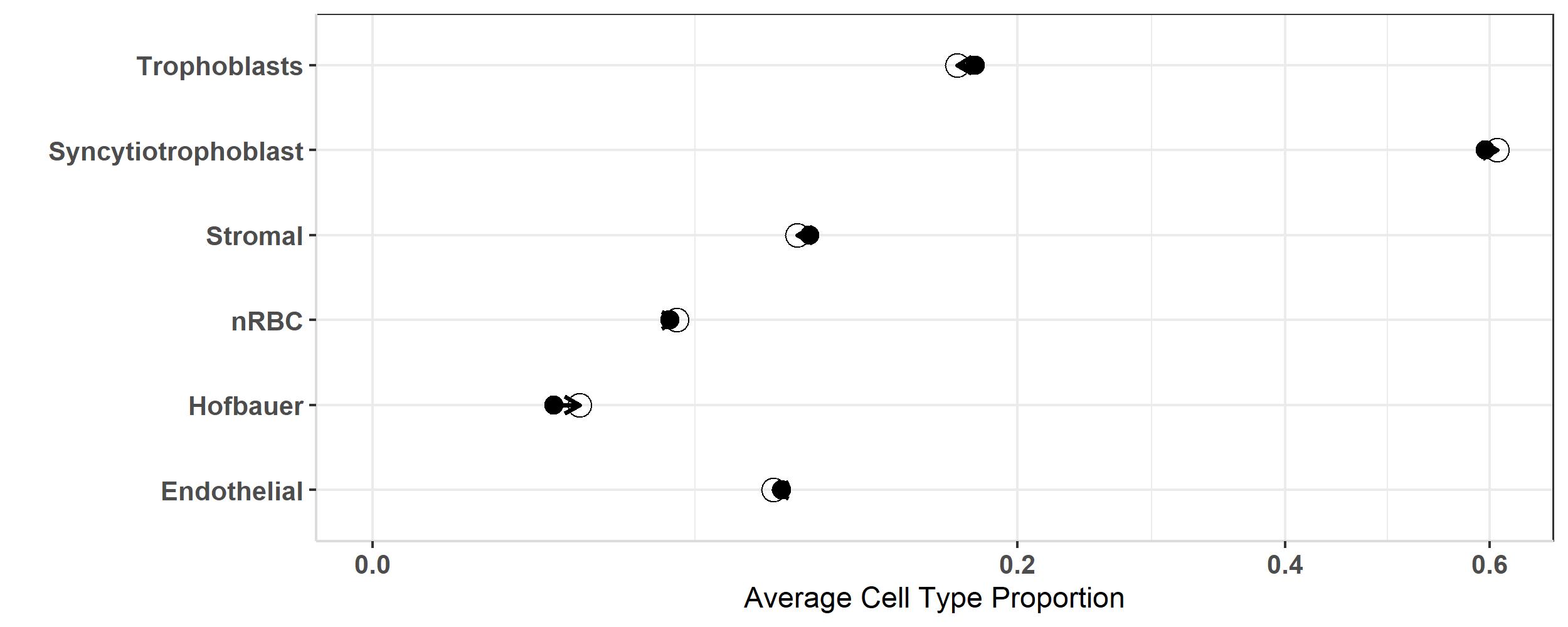}
\caption{Visual summary of the DBWQS regression results showing the log-transformed estimated absolute change (X-axis) in each cell type proportion (Y-axis) for a quartile increase in placental chemical mixture index. Dark circles represent the average cell proportion observed in the study population; white circles represent the estimated cell proportion associated with a quartile increase in placental chemical mixture index. Arrows indicate the direction of the change. } 
\label{fig:changeplot-richs}
\end{figure}

The effect estimates of the DBWQS regression, linking the chemical mixture to cell proportions, can be interpreted as the expected percent change in absolute proportion of each cell type for a one-unit (i.e. quartile) increase in the chemical mixture. For each cell type and chemical, we reported the effect estimates (in percentage) along with the $95\%$ credible intervals ($95\%$ CrI), effective sample size (ESS) and $\hat{R}$. All results are summarized in \textbf{Table~\ref{tbl-richs-res}} and in \textbf{Figures~\ref{fig:barplot-mix-weights} \&~\ref{fig:changeplot-richs}}.

The results showed that a quartile increase in the DBWQS mixture index was associated with a $30.4\%$ ($95\%$ CrI: $[0.8\%, 71.1\%]$) increase in Hofbauer cell proportion, a (suggesting) decrease in Trophoblasts cell proportion $-6\%$ ($80\%$ CrI: $[-10.7\%, -1.4\%]$) and a (suggesting) increase in Syncytiotrophoblasts cell proportion $2.2\%$ ($80\%$ CrI: $[0.4\%, 4\%]$) . Among the seven chemicals (\textbf{Figure~\ref{fig:barplot-mix-weights}}), As was the major driver (bad actor) of the mixture association with an estimated weight of $0.26$, followed by Se and Cr with estimated weights of $0.20$ and $0.19$, respectively. \textbf{Figure~\ref{fig:changeplot-richs}} provides a visual summary of the direction and magnitude of change in absolute cell type proportions associated with a quartile increase of the DBWQS mixture.
For completeness, results of the change of cell proportion relative to syncytiotrophoblast are reported in the Supplemental material (\textbf{Table~\ref{tbl-richs-res-relprop}}).

\begin{table}[H]
\centering
\begin{tabular}{ |c|c|c|c|c|c| } 
 \hline
 \textbf{Cell Types} & \textbf{Effect Estimate} & $\pmb{95\%}$\textbf{ CrI} & $\pmb{80\%}$\textbf{ CrI} & \textbf{ESS}  & $\hat{\textbf{R}}$ \\
 \hline
 Syncytiotrophoblasts & 2.2\% & [-0.7\%, 5.3\%] & [0.4\%, 4\%] & 8865 & 1.000 \\
 Trophoblasts & -6\% & [-13.7\% , 1.2\%] & [-10.7\%, -1.4\%] & 13038 & 1.000 \\
 Stromal & -5.6\% & [-17.3\% , 4.7\%] & [-12.5\%, 1.1\%] & 11345 & 1.000 \\
 \textbf{Hofbauer} & \textbf{30.4\%} & \textbf{[0.8\% , 71.1\%]} & \textbf{[9.8\%, 53.3\%]} & 12750 & 1.000 \\
 Endothelial & -3.8\% & [-14\% , 8\%] & [-10.5\%, 3.3\%] & 13489 & 0.999 \\
 nRBC & 4.4\% & [-10\% , 22\%] & [-5.3\%, 14.6\%] & 14877 & 0.999 \\
 \hline
 \hline
 \textbf{  Chemicals} & \textbf{Weight Estimate} & $\pmb{95\%}$\textbf{ CrI} & $\pmb{80\%}$\textbf{ CrI} & \textbf{ESS}  & $\hat{\textbf{R}}$ \\
 \hline
 Arsenic & 0.26 & [0 , 0.68] & [0.002, 0.21] & 7551 & 1.000 \\
 Selenium & 0.20 & [0, 0.51] & [0.003, 0.26] & 9772 & 1.000 \\
 Chromium  & 0.19 & [0, 0.50] & [0.02, 0.38] & 1618 & 1.001 \\
 Cobalt & 0.11 & [0, 0.37] & [0, 0.12] & 11292 & 0.999 \\
 Lead  & 0.10 & [0, 0.35] & [0.001, 0.24] & 10477 & 1.000 \\
 Aluminum & 0.09 & [0, 0.30] & [0.02, 0.40] & 10075 & 0.999 \\
 Nickel & 0.05 & [0, 0.21] & [0.01, 0.54] & 13432 & 0.999 \\
 \hline
\end{tabular}
\caption{(Effect and Weight) Estimates, $95\%$ and $80\%$ credible intervals (CrI), effective sample size (ESS) and $\hat{R}$ of the percentage change of the absolute cell type proportions for one quartile increase in the chemical mixture and the weights of the seven chemicals. $\hat{R}$ helps with chain convergence diagnosis.} 
\label{tbl-richs-res}
\end{table}

\section{Discussion}\label{sec:disc}
This is the first study developing an environmental mixture approach for compositional outcomes. We showed that the DBWQS regression accurately assesses associations between environmental exposure mixtures and compositional outcomes. DBWQS jointly models compositions as a multivariate outcome using the alternative parameterization of the Dirichlet distribution, while handling the exposure mixture as predictor. We derived the posterior distribution of our model, and developed R-Stan code to perform Bayesian inference via Hamiltonian Monte Carlo. We reported the performance and parameter estimates of DBWQS regressions on 72 simulated scenarios (each with 100 Monte Carlo repetitions), as well as a real case scenario. All simulated and observational data are available, together with code on (\textit{github.com/hasdk/xbwqs}).
The DBWQS approach achieves low RMSE for all parameter estimates and good convergence rates across simulated data, although it performs better with datasets of larger sample size and smaller number of exposures. Notably, we showed that jointly analyzing the compositional outcomes using DBWQS yielded significantly improved performance (lower bias, RMSE and standard errors, and more accurate coverage probabilities) compared with modeling each outcome proportion independently. 

In the epigenetic application, we leveraged publicly available data from the RICHS cohort to estimate the association between DNA methylation derived proportions of six placenta cell types and a chemical mixture exposure, while adjusting for independent covariates able to confound the association. We showed that a quartile increase in chemical mixture was associated with a significant increase in proportions of Hofbauer cells, with As, Se and Cr driving most of the mixture effect. Our findings corroborate previous environmental health studies investigating placenta cells, including Hofbauer, in association with cadmium and arsenic levels. Hofbauer cells in the placenta were influenced by increased cadium levels in cord blood \citep{bulka2022}; and \cite{priya2011} found a significant increase in the number Hofbauer cells among preeclamptic placentas compared to normal controls. These findings suggest altered placental function and potential change in immune status as a results of increased exposure to toxic environmental exposures. 

The current DBWQS model has a single parameter ($\phi$) that controls the variance of all the components of the compositional outcome. Future method developments may consider linking the variance parameter to the independent variables (covariates) using an exponential link function in order to better capture the variability in the outcomes, but that will come at the cost of increased model complexity. Another consideration is to estimate separate mixture weights ($\pmb{w}$) for each component of the compositional outcome instead of the current implementation considering a single vector of mixture weights for the entire compositional outcome. Our rationale for modeling a single weights vector for all categories was mainly motivated by our desire to keep a tractable number of parameters so that the method can scale well to epidemiologic studies where the number of exposures and outcomes are moderate to large relative to the sample sizes. 

In addition, both the Dirichlet and DBWQS regression models lack flexibility in modeling the simplical dependencies between the outcomes \citep{ascari_multivariate_2023}. In future work, we plan to improve the flexibility of the DBWQS by exploring some of the recent generalizations of the Dirichlet distribution, such as the extended flexible Dirichlet (EFD) which is a mixture of Dirichlet components \citep{ongaro2020}. 

In summary, this approach provides a solution in environmental health to studies aiming to analyze associations between mixtures of chemical exposures and health outcomes that are compositional in nature. Examples of such outcomes go beyond cell type proportions explored in our real data application, and can include microbiome data, daily time allocation of sleep and physical activity, and daily consumption of different nutrient groups (proteins, carbohydrates, fats, and micro-nutrients).

\newpage
\bibliographystyle{apalike}
\bibliography{bwqs-comp-refs}

\newpage
\section{Supplementary Materials}

\subsection{Deriving the posterior distribution for DBWQS regression model}

Recall that the DBWQS regression model for i.i.d observations $\{\pmb{y}_i, \pmb{x}_i, \pmb{q}_i\}_{i=1,...,N}$, the regression model can be written as
\begin{align}
    p(\pmb{y}_i | \pmb{x}_i, \pmb{q}_i) &\sim Dirichlet(\pmb{\mu}_i, \phi), \\
    \mu_{i,k} &= \frac{\exp(S_i \theta_k  + \pmb{x}_i^T \pmb{\beta}_{k})}{\sum_{c=1}^K \exp(S_i\theta_c + \pmb{x}_i^T \pmb{\beta}_{c})} , \ \text{for} \ k=2,...,K,\\
    \mu_{i,1} &= \frac{\exp(S_i . 0 + \pmb{x}_i^T . 0)}{1 + \sum_{c=2}^K \exp(S_i \theta_c + \pmb{x}_i^T \pmb{\beta}_{c})} = \frac{1}{1 + \sum_{c=2}^K \exp(S_i \theta_c +\pmb{x}_i^T \pmb{\beta}_{c})},\\
    S_i &= \sum_{m=1}^M q_{i,m} w_m,\\
    \pmb{\beta}_1 &= \pmb{0} \ \textit{and} \ \theta_1 = 0 \quad \ \textit{(base or reference category)},\\
    \pmb{w} &\sim Dirichlet(\pi_1, ..., \pi_M), \\
    \pi_m &\sim Gamma(2,2), \quad \text{for} \ m=1,...,M,\\
    \beta_{k,j} &\sim Normal(0, \sigma_{\beta}), \quad \text{for} \ k=2,...,K, and \ j=1,...,J, \\
    \theta_k &\sim Normal(0, \sigma_{\theta}), \quad \text{for} \ k=2, ..., K, \\
    \phi &\sim Gamma(a,b), \quad \text{for} \ a>0, \ b>0
\end{align}

The posterior distribution under the compositional BWQS regression model with multinomial-logit link can be written as
\begin{align}
p(\pmb{\beta}, \pmb{w}, \pmb{\pi}, \pmb{\theta}, \phi | \pmb{y}, \pmb{x}, \pmb{q}) &\propto p(\pmb{y} | \pmb{x}, \pmb{q}; \pmb{\beta}, \pmb{w}, \pmb{\pi}, \pmb{\theta}, \phi) . p(\pmb{w}|\pmb{\pi}) . p(\pmb{\pi}) . p(\pmb{\beta}) . p(\pmb{\theta}) . p(\phi)\\
&= \prod_{i=1}^N \left[ \frac{\Gamma(\phi_i)}{\prod_{k=1}^K \Gamma(\phi_i . \mu_{i,k})}  \prod_{k=1}^K y_{i,k}^{(\phi_i . \mu_{i,k})-1} \right] . \left[ \frac{\Gamma(\sum_{m=1}^M \pi_m)}{\prod_{m=1}^M \Gamma(\pi_m)} \ \prod_{m=1}^M w_m^{\pi_m - 1} \right] \nonumber \\
&\quad \quad . \left[ \prod_{m=1}^M  \ \frac{\pi_m}{4} . \exp\left(-\frac{\pi_m}{2}\right) \right]. \left[ \left( \sigma_{\beta} \sqrt{2\pi}\right)^{-KJ}\prod_{k=1}^K \prod_{j=1}^J \exp \left(-\frac{\beta_{k,j}^2}{2\sigma_{\beta}^2} \right) \right] \nonumber \\
&\quad \quad . \left[ \left( \sigma_{\theta} \sqrt{2\pi}\right)^{-K} \prod_{k=1}^K \exp\left(-\frac{\theta_{k}^2}{2\sigma_{\theta}^2}\right) \right] . \left[ \frac{b^a}{\Gamma(a)} \phi^{a-1} \exp(-b.\phi) \right]
\end{align}

\subsection{Real Data Application}
\paragraph{Association results in relative proportions}

Table~\ref{tbl-richs-res-relprop} summarizes the coefficients estimates from DBWQS on the RICHS cohort in relative proportions scale. We used Syncytiotrophoblast cells as our reference cell type since they are the most abundant; therefore, the coefficients in the table are interpreted as changes in proportions of each cell type relative to the proportion of Syncytiotrophoblast cells.

\begin{table}[H]
\centering
\begin{tabular}{ |c|c|c|c|c|c| } 
 \hline
 \textbf{Cell Types} & \textbf{Effect Estimate} & $\pmb{95\%}$\textbf{ CrI} & $\pmb{80\%}$\textbf{ CrI} & \textbf{ESS}  & $\hat{\textbf{R}}$ \\
 \hline
 Trophoblasts & -0.08 & [-0.19, 0.01] & [-0.15, -0.02] & 11142 & 1.000 \\
 Stromal & -0.08 & [-0.22 , 0.04] & [-0.16, -0.002] & 10217 & 1.000 \\
 \textbf{Hofbauer} & \textbf{0.23} & [-0.02, 0.52] & \textbf{[0.07, 0.41]} & 13196 & 1.000 \\
 Endothelial & -0.06 & [-0.19, 0.07] & [-0.14, 0.02] & 12114 & 0.999 \\
 nRBC & 0.02 & [-0.14, 0.19] & [-0.08, 0.12] & 14111 & 0.999 \\
 \hline
\end{tabular}
\caption{Estimates, $95\%$ and $80\%$ credible intervals (CrI), effective sample size (ESS) and $\hat{R}$ of the change in the cell type proportions relative to Syncytiotrophoblasts (the reference cells), for one quartile increase in chemical mixture. $\hat{R}$ is used to evaluate chain convergence.} 
\label{tbl-richs-res-relprop}
\end{table}

\subsubsection{Trace Plots}
\begin{figure}[!htb]
\centering
\includegraphics[width=\textwidth]{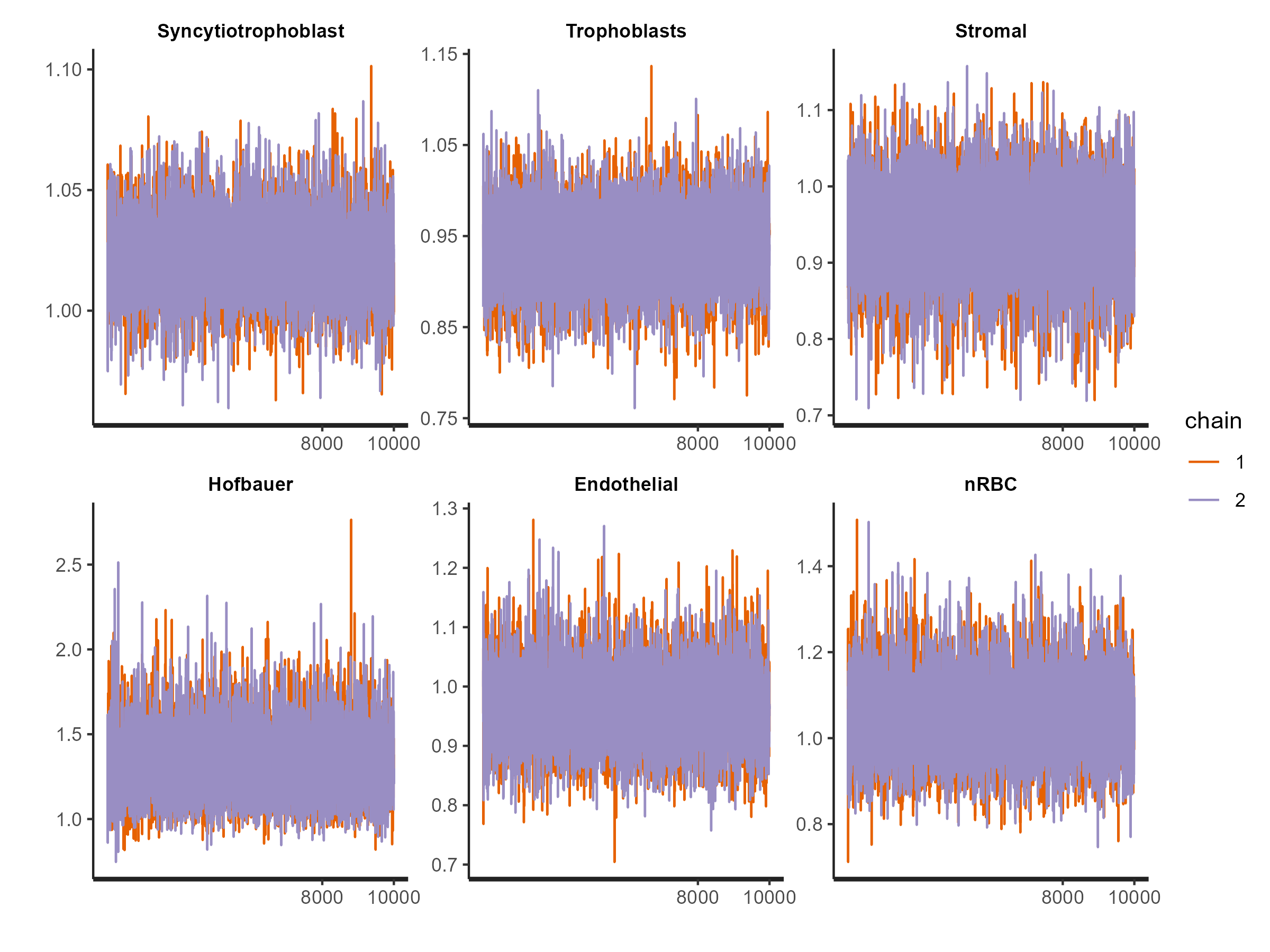}
\caption{Trace plot for estimated coefficients (in absolute proportion scale) from DBWQS model on RICHS data application.} 
\label{fig:traceplot_absprop}
\end{figure}

\begin{figure}[!htb]
\centering
\includegraphics[width=\textwidth]{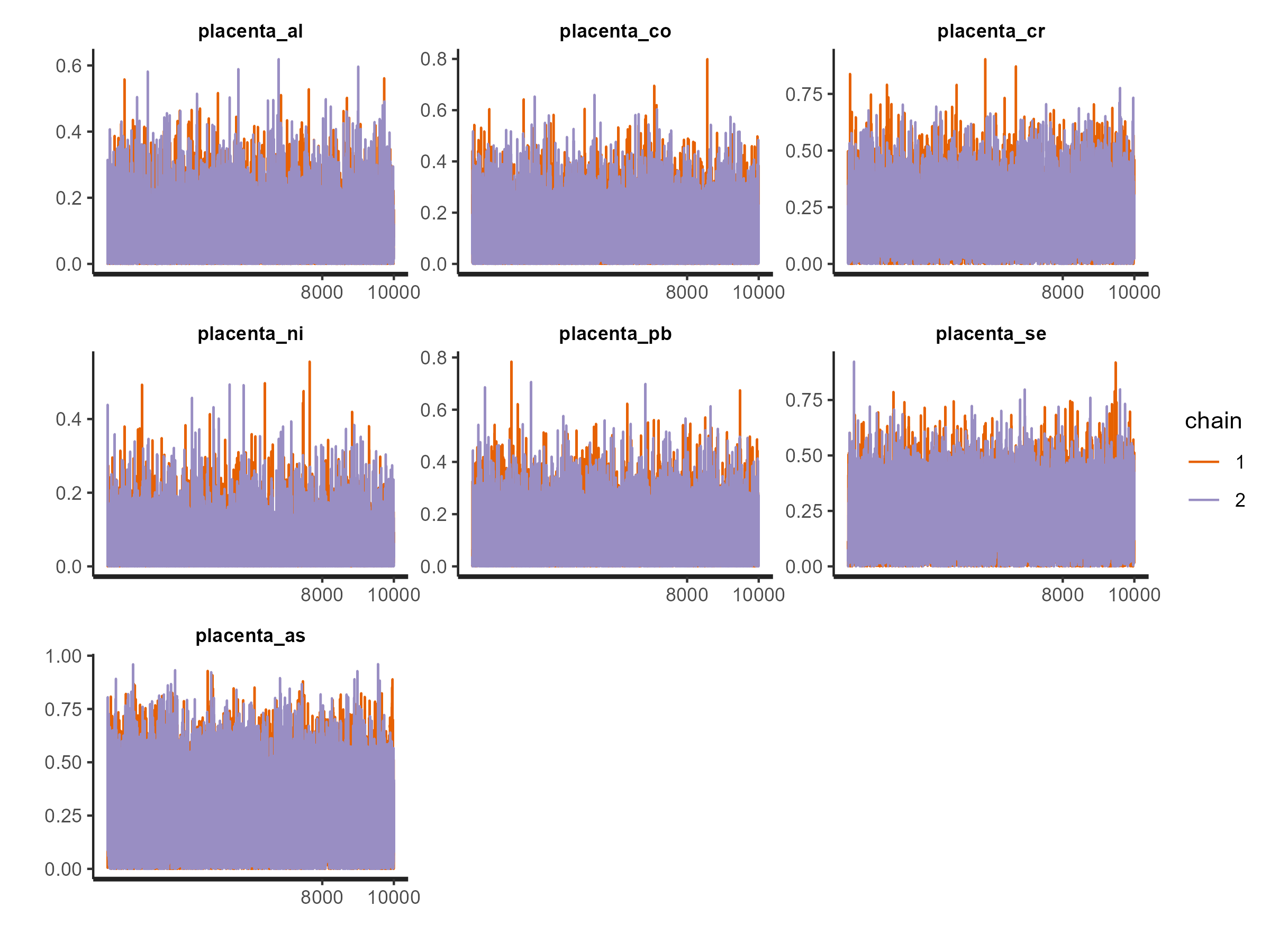}
\caption{Trace plot for estimated mixture weights from DBWQS model on RICHS data application.} 
\label{fig:traceplot_mixwt}
\end{figure}

\subsubsection{Auto-correlation Plots}
\begin{figure}[!htb]
\centering
\includegraphics[width=\textwidth]{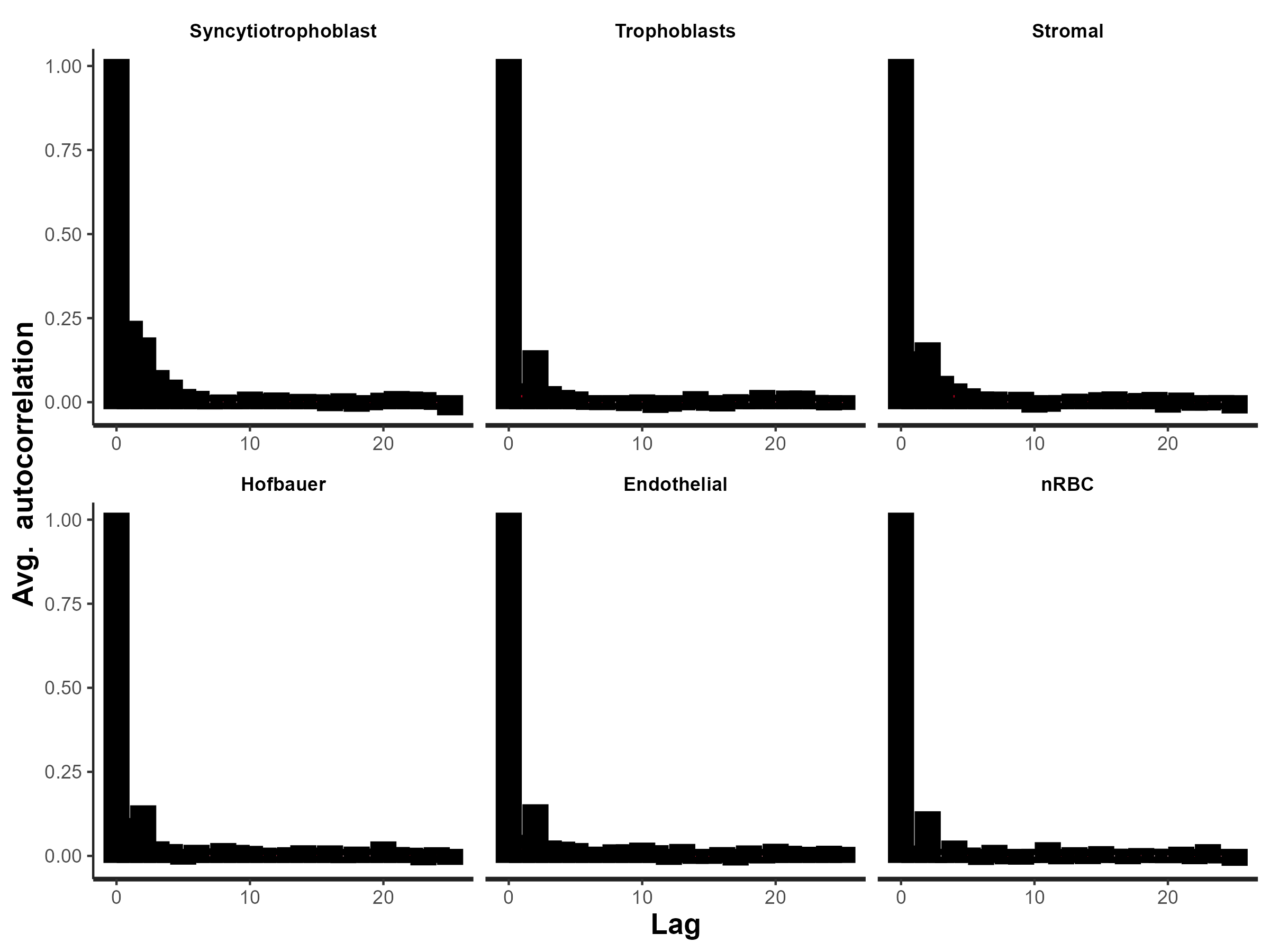}
\caption{Auto-correlation plot for estimated coefficients (in absolute proportion scale) from DBWQS model on RICHS data application.} 
\label{fig:acplot_abscoef}
\end{figure}

\begin{figure}[!htb]
\centering
\includegraphics[width=\textwidth]{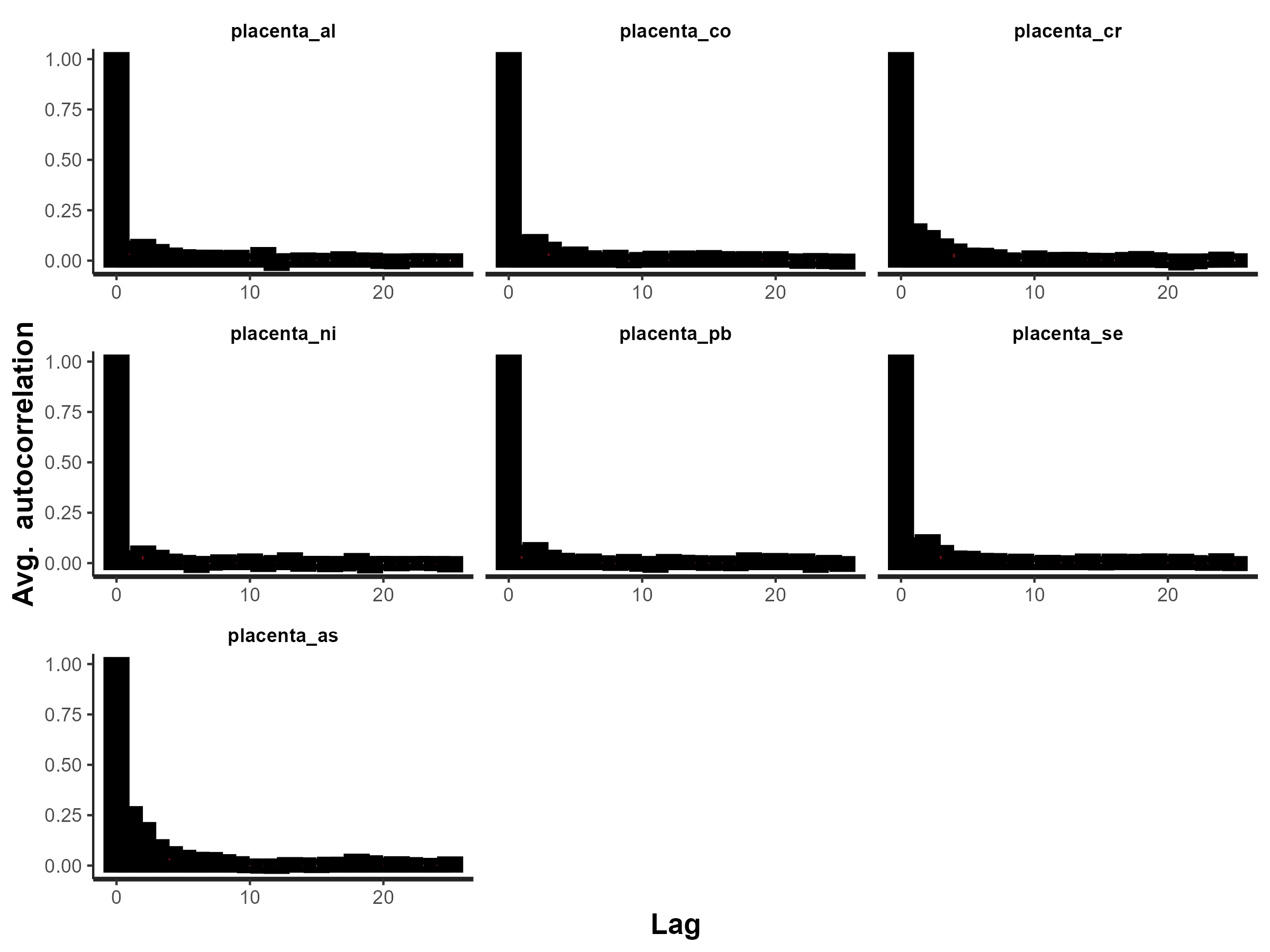}
\caption{Auto-correlation plot for estimated mixture weights from DBWQS model on RICHS data application.} 
\label{fig:acplot_mixwt}
\end{figure}

\end{document}